\documentclass[aps,pre,floatfix,nofootinbib,showpacs,twocolumn]{revtex4-1}
\usepackage{amsmath}    
\usepackage{graphicx}
\usepackage{amssymb}
\begin{document}
\author{George Thomas\footnote{electronic address: george@iisermohali.ac.in}  
and Ramandeep S. Johal\footnote{electronic address: rsjohal@iisermohali.ac.in}}
\affiliation{Indian Institute of Science Education and Research Mohali,\\
Transit Campus: MGSIPAP Complex, Sector 26, Chandigarh 160019, India}
\draft
\title{A Coupled Quantum Otto Cycle}
\begin{abstract}
We study the 1-d isotropic Heisenberg model of two spin-1/2
systems as a quantum heat engine. The engine undergoes a four-step Otto cycle
where 
the two adiabatic branches involve changing the  external magnetic field at a
fixed 
value of the coupling constant. We find conditions for the engine 
efficiency to be higher than the uncoupled model; in particular, we find an 
upper bound which is tighter than the Carnot bound. A new domain of parameter 
values is pointed out  which was not feasible in 
the interaction-free model. Locally, each spin seems to
effect the flow of heat in a direction opposite to the global temperature
gradient. 
This seeming contradiction to the second law can be resolved in terms of local
effective temperature of the spins. 
 \end{abstract}
\maketitle
\section*{I. Introduction}
Quantum generalisations of classical heat cycles have now been studied for some 
years. When the working medium is a few-level quantum system, new lines of
enquiry
open up due to additional features like discreteness of states, 
quantum correlations, quantum coherence and so on 
\cite{Scovil,Kieu2004,TZhang,GFZhang,Wang2009,Scully,Lutz}. Many models have
served to 
investigate the validity of second law of thermodynamics in the quantum
regime\cite{Leff,Quan2006}.
The possibility of small scale devices and information processing machines \cite{Zhou2010} has
generated further interest into the fundamental limits imposed
on the heat generation,
cooling power and thermal efficiencies achievable with these models
\cite{Allahv2010,Noah2010,Paul}. 
Quantum analogues of Carnot cycles, Otto cycles and other brownian machines
have been analysed \cite{Quan2007,Humphrey2002}. Further, both infinite
\cite{Kieu2004,TZhang,GFZhang} and 
finite-time
\cite{Geva1992,Broeck2005,Feng,AJM2008,Esposito2009,ELB2009,EKLB2010}
thermodynamic cycles have attracted attention.

The quantum Otto cycle which occupies our interest here consists of a working
substance
with hamiltonian $H$ and initial density matrix $\rho$  being manipulated
between two heat reservoirs (the reservoir temperatures satisfy $T_1 > T_2$) 
under two adiabatic and two isochoric branches. 
On the adiabatic branches, the system is assumed to follow quantum adiabatic
theorem
and thermodynamic work is defined in terms of the change in energy levels at
given occupation probabilities. If the hamiltonian is changed from $H_1$ to
$H_2$ by controlling
an external parameter then the work performed is defined as ${\rm
Tr}[\rho(H_2-H_1)]$.
On the other hand, while traversing the isochoric branches, heat is exchanged
with the reservoirs. Thus if the density matrix of the system changes
from $\rho_1$ to $\rho_2$ for a given hamiltonian $H$, then heat exchanged is 
${\rm Tr}[(\rho_2-\rho_1)H]$.  As an example, for an effectively
two-level system whose energy splitting can be varied from $E_1$ to $E_2$, 
the Otto efficiency has been found to be $1-E_2/E_1$, which is bounded from
above
by Carnot value due to the condition $E_2/E_1 >T_2/T_1$ \cite{Kieu2004}. 

Recently, some authors have studied the role of different quantum interactions 
using spin-1/2 particles in a Quantum Otto cycle 
\cite{TZhang,GFZhang,Wang2009}. 
In particular, the role
of quantum entanglement has been conjectured using measure like concurence
and the second law has been shown to hold in such models. 
In this paper, we also investigate a coupled Otto engine using
a 1-d Heisenberg model with isotropic exchange interactions
between two spin-1/2 particles (see Eq. (\ref{H}) below). In 
\cite{TZhang} the same model was analysed, where during the adiabatic steps,
the exchange constant $J$ was altered between two chosen values ($J_1 \to J_2
\to J_1$),
while keeping the external magnetic field at a fixed value. 
From an experimental point of view, it is also interesting
to investigate a cycle where the exchange constant is fixed 
 and only the magnetic field
is varied during the adiabatic steps. Further, 
the uncoupled model cycles  considered earlier in literature
can be taken as a benchmark with which to compare the engine performance of the 
coupled model. 

The paper is organised as follows. In section II, we present the quantum model
of our working medium, enumerating the energy eigenstates and eigenvalues.
In IIA, the various stages of the heat cycle are described and expressions 
for heat exchanged with reservoirs and work delivered are calculated.
It is instructive to develop the engine operation based on local description.
It is shown that all the work is done locally by each spin.  In subsections IIIA
and IIIB we develop two cases i) $B_1 > B_2$ and  ii) $B_2 > B_1$. The latter case 
is possible only in the presence of interactions. It is observed for this 
case that second law of thermodynamics can be violated at the local scale. 
General conditions are derived
when the efficiency is higher than the noninteracting model. We also 
present an upper bound for efficiency which is lower than the Carnot bound.
The proof is sketched in the Appendix. In IIIC, we interpret some nontrivial
features of the engine operation in terms of local spin temperatures. 
The final section IV summarises our findings.
\section*{II. The coupled QHE}
The working medium for our QHE  consists of 
two spin-1/2 particles within the 1D isotropic Heisenberg 
model \cite{TZhang,Arnesen}. The Hamiltonian is given by
\begin{equation}
H = J({\sigma^1}.{\sigma}^2 + {\sigma}^2.{\sigma}^1)+
B(\sigma^1_z+\sigma^2_z),
\label{H}
\end{equation}
where ${\sigma^{1(2)}}=(\sigma^{1(2)}_{x},\sigma_{y}^{1(2)} ,\sigma_{z}^{1(2)})$
are the Pauli matrices, $J=J_x =J_y=J_z$ is the exchange constant and $B$ is the
magnetic field along $z$-axis. Cases $J>0$ and $J<0$ correspond to 
antiferromagnetic and ferromagnetic interactions, respectively. In this paper,
we consider antiferromagnetic 
case only. The  energy eigenvalues of ${H}$ are  
$ -6J$, $(2J-2B)$, $2J$ and $(2J + 2B)$. If $\arrowvert0\rangle$ and 
$\arrowvert1\rangle$ represent the state of the  spin
along and opposite to the direction of the magnetic field respectively, then  in
the natural basis
$\{ \arrowvert11\rangle$, $\arrowvert10\rangle$, $\arrowvert01\rangle$,
$\arrowvert00\rangle \}$, we can write the density matrix as 
\begin{equation}
 \rho=P_1\arrowvert\psi_-\rangle\langle\psi_-|+P_2\arrowvert00\rangle\langle00|+
P_3\arrowvert\psi_+\rangle\langle\psi_+|+P_4\arrowvert11\rangle\langle11|,
\end{equation}
where $|\psi_\pm \rangle = (\arrowvert10\rangle\pm\arrowvert01\rangle)/\sqrt{2}$
are the maximally entangled Bell states. The occupation probabilities of the
system in
the thermal state at temperature $T$ are given by
\begin{eqnarray}
\*P_1&=&\frac{e^{8J/T}}{Z}\\
P_2&=&\frac{e^{2B/T}}{Z},\\
P_3&=&\frac{1}{Z},\\
P_4&=&\frac{e^{-2B/T}}{Z}.
\end{eqnarray}
where, $Z=(1+e^{8J/T}+e^{2B/T}+e^{-2B/T})$ is the normalisation constant.
\begin{figure}
\includegraphics{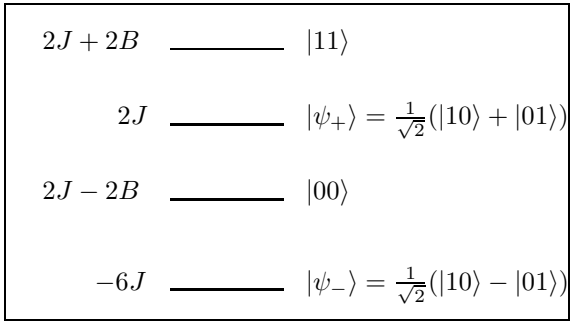}
\caption{Energy eigenvalues and eigenstates of two-spins entangled model system}
\end{figure}
%
\subsection*{A. The heat cycle}
The four stages invloved in our quantum Otto cycle are described below: \\
\textit{Stage 1}:
the  system with the external magnetic field at $B_1$ attains thermal
equilibrium
with a bath of 
temperature $T_1$. Let occupation probabilities be $p_1$, $p_2$, $p_3$, and
$p_4$ as tabulated above with $T=T_1$ and $B= B_1$.
\textit{Stage 2}:
the system is isolated from the hot bath and  the magnetic field is changed 
from $B_1$ to $B_2$ by an adiabatic process. According to quantum adiabatic
theorem, the process should be  slow  enough to maintain the individual
occupation probability of each energy level.
\textit{Stage 3}:
the system is brought in thermal contact with a cold bath at temperature $T_2$.
Upon attaining equilibrium with the bath, the occupation probabilities become
$p'_1$, $p'_2$, $p'_3$, and $p'_4$ corresponding to the thermal state with
$T=T_2$ and $B= B_2$.
On the average, the system gives off heat to the bath.
\textit{Stage 4}:
the system is removed from the cold bath and undergoes another quantum adiabatic
process which changes the magnetic field from $B_2$ to $B_1$ but keeps the 
probabilities $p_1'$, $p_2'$, $p_3'$, and $p_4'$ unaffected.
 Finally, the system is brought back to touch the hot bath.
On the average, heat is absorbed from the bath and the system returns to its
initial state.

The heat transfered in \textit{Stage 1} and in \textit{Stage 3} of the cycle
respectively is
\begin{eqnarray}
Q_1&=&\sum_{i} E_i(p_i-p_i') \\ 
 & = & 8J(p_1'-p_1)+2B_1(p_2'-p_2+p_4-p_4'),
\label{q1}
\end{eqnarray}
and
\begin{eqnarray}
Q_2&=&\sum_{i} E_i'(p_i'-p_i) \\
& = & -8J(p_1'-p_1)-2B_2(p_2'-p_2+p_4-p_4').
\label{q2}
\end{eqnarray}
In the above, $E_i$ and $E_i'$ ($i=1, 2, 3, 4$) are the energy eigenvalues of
the system in
 \textit{Stage 1} and \textit{Stage 3} respectively. $Q_1>0$ and
$Q_2<0$ corresponds to absorption of heat from hot bath and release of heat to
cold bath respectively. Comparing the equations for heat transfer between the
system
and the reservoirs, Eqs. (\ref{q1}) and  (\ref{q2}), the quantity of heat  
$8J(p_1'-p_1)$  appears in both the equations. Obviously, this term is absent
in the uncoupled case for which $J=0$.
As will be shown below, the sign ($\pm$) of this term
determines whether the efficiency in the coupled case will be higher or lower 
than the uncoupled case.

The work is  done in \textit{Stage 2} and \textit{Stage
4} when the energy levels are changed at fixed occupation probabilities.
The net work done by the QHE is
\begin{equation}
W=Q_1+Q_2=2(B_1-B_2)(p_2'-p_2+p_4-p_4').
\label{4}
\end{equation}
Note that $W>0$ corresponds to work performed by the system. 
\section*{III. The local description}
 In this section, we discuss how the individual spins in the system undergo the
cycle.
Again, let $\varrho_{12}$ and $\varrho_{12}'$ represent the thermal states in
the natural basis 
 when the system  is in equilibrium in \textit{Stage 1} and
\textit{Stage 3} respectively. Explicitly, the density matrices are
\begin{equation}
\varrho_{12}=\left(\begin{array}{cccc}
p_4&0&0&0\\
0&\frac{p_1+p_3}{2}&\frac{p_3-p_1}{2}&0\\
0&\frac{p_3-p_1}{2}&\frac{p_1+p_3}{2}&0\\
0&0&0&p_2 \end{array}\right),
\end{equation}
\begin{equation}
\varrho_{12}'=\left(\begin{array}{cccc}
p_4'&0&0&0\\
0&\frac{p_1'+p_3'}{2}&\frac{p_3'-p_1'}{2}&0\\
0&\frac{p_3'-p_1'}{2}&\frac{p_1'+p_3'}{2}&0\\
0&0&0&p_2' \end{array}\right).
\end{equation}
Let 
$\varrho_{1}$ and $\varrho_{2}$ be the reduced density matrices in 
\textit{Stage 1} for the first and the second spin, respectively. 
Then from the normalization constraints,
$\Sigma_{i}^{} p_i=\Sigma_{i}^{} p_i'=1$, we get 
\begin{equation}
\varrho_{1}=\varrho_{2}=\left(\begin{array}{cc}
\frac{1}{2}-\frac{(p_2-p_4)}{2} & 0\\
0&\frac{1}{2}+\frac{(p_2-p_4)}{2} \end{array}\right).
\label{rho}
\end{equation}
Similarly in \textit{Stage 3}, the reduced density matrices for the first and
second spin are 
\begin{equation}
\varrho_{1}'=\varrho_{2}'=\left(\begin{array}{cc}
\frac{1}{2}-\frac{(p_2'-p_4')}{2} & 0\\
0&\frac{1}{2} + \frac{(p_2'-p_4')}{2} \end{array}\right).
\label{rho_prime}
\end{equation}
Since the applied magnetic field  is the same for each spin, their local
Hamiltonian
is also same. Let $H_{l}$ and $H_{l}'$ be the  local Hamiltonians for
individual spins  with eigenvalues $(B_1, -B_1)$ and $(B_2, -B_2)$ in
\textit{Stage 1}
and \textit{Stage 3} respectively.
The heat transferred locally between \textit{one} spin and a reservoir 
is given by  
\begin{eqnarray}
q_1&=&B_1(p_2'-p_2+p_4-p_4'),\\
q_2&=&-B_2(p_2'-p_2+p_4-p_4'), 
\end{eqnarray}
for the hot and the cold reservoir, respectively.
So we get the net work done by an individual spin as 
\begin{equation}
w=q_1+q_2=(B_1-B_2)(p_2'-p_2+p_4-p_4').
\label{3}
\end{equation}
From Eqs. (\ref{3}) and (\ref{4}) 
\begin{equation}
W=2w.
\end{equation}
Thus the total work performed is the sum of work obtained from the two spins
locally. 

Further, the total heat absorbed by the system can be written
as
\begin{equation}
Q_1 =  8J(p_1'-p_1) + 2 q_1,
\end{equation}
and similarly for the heat released to the cold bath is
\begin{equation}
Q_2 = - 8J(p_1'-p_1) + 2 q_2.
\end{equation}
Now it can be seen that because the work is done only due to
change in local hamiltonians, so only the part of the heat 
which is absorbed locally by a spin can be converted into heat.
The part $8J(p_1'-p_1)$ cannot potentially be converted into work 
due to the nature of the adiabtic process involved and
is transferred directly between the reservoirs. But it may not be
transfered only  from the hot to the cold bath, in which case it may be regarded
like
a heat leakage term. In fact, the flow of this heat can be in the opposite 
direction which is directly related to the enhancement of efficiency due
 to coupling, as shown below.

In the following, we consider two cases whereby magnetic field may be
decreased or alternately, increased in Stage 2. It will be seen that the second
case is 
feasible only in the presence of interactions, $J\ne 0$. In the first case when
$J = 0$,
the above equations go back to those for Kieu's model with two
uncoupled spins where an engine operation is obtained given $T_1 >T_2$
and $B_1 > B_2$ with the additional condition $B_2/T_2 > B_1/T_1$.
%
\subsection*{A. The case $B_1> B_2$} 
From Eq. (\ref{4}), the condition that the work performed be positive ($W>0$)
is given by
\begin{equation}
(p_2' - p_4') > (p_2- p_4).
\label{pow}
\end{equation}
Secondly, for the heat to be absorbed from the hot bath 
($Q_1 > 0$), from Eq. (\ref{q1}) we have one of the following two possibilities:
(i) $p_1'> p_1$ or (ii) $p_1'< p_1$. Alongwith 
the possibility (ii), we must also have
$(p_2' - p_2  + p_4 -p_4') > (4 J/B_1) ({p_1} - p_1')$.
Now we rewrite Eq. (\ref{q1}) as 
\begin{equation}
Q_1 = 8 J (p_1' - p_1) + \frac{W B_1}{(B_1-B_2)},  
\end{equation}
or $8 J (p_1' - {p_1}) = Q_1 (1-\eta/\eta_0)$,
where $\eta = W/Q_1$ is the efficiency
of the coupled engine and $\eta_0 = (B_1-B_2)/B_1$ is the efficiency of the
uncoupled i.e. $J=0$ case. Thus for $J> 0$, if $p_1'> p_1$, then $\eta <
\eta_0$,
or the presence of  coupling between the spins decreases the efficiency 
from its value $\eta_0$.  The global efficiency is equal
to the 
local efficiency in two situations, when $J=0$ or $p_1=p_1'$.

\begin{figure}[ht]
\includegraphics[width=8cm]{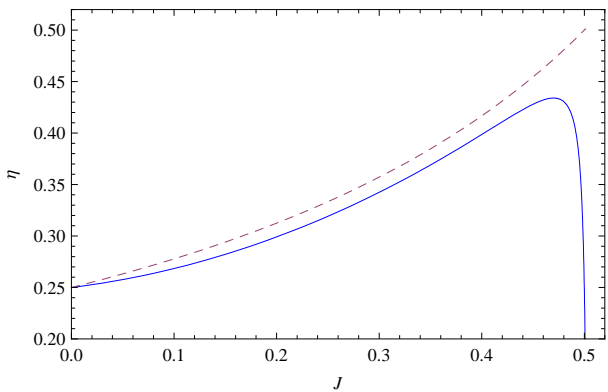}
\caption{Efficiency versus the coupling constant $J$, for $B_1 > B_2$ case, for
values 
$B_1=4$, $B_2=3$, $T_1=1$ and $T_2=0.5$. The uncoupled case corresponds to
$\eta_0 = 1-B_2/B_1 = 0.25$.
 The upper curve denotes the bound for efficiency from Eq. (\ref{etabound}).
Beyond the maximum
shown value of $J$, the efficiency drops monotonically. The Carnot limit is
0.5.}
\label{fig2}
\end{figure}

On the other hand,
if $p_1'< p_1$, then it is possible that the efficiency of the coupled
engine can be higher than the uncoupled case. Using the latter condition with
Eq. (\ref{pow}), 
we have 
\begin{equation}
\frac{(p_2' - p_4')}{p_1'} > \frac{(p_2- p_4)}{p_1}.
\end{equation}
From the explicit expressions for the probabilities, the above inequality can 
be simplified to give 
\begin{equation}
\frac{B_2}{T_2}  > \frac{B_1}{T_1}. 
\label{bt}
\end{equation}
Thus we see that the above condition which is necessary to extract work
in the $J=0$  model is \textit{also} the  condition for the coupled case to
obtain an efficiency 
higher than  $\eta_0$.
But additionally, for a set of given values of $T_1$, $T_2$, $B_1$ and $B_2$,
there is a
maximum 
value of $J$ beyond which  
the efficiency drops below the $\eta_0$ value.  See Fig. \ref{fig2}.

The reason for the lowering of efficiency when $p_1<p_1'$, is that the term
$8J(p_1'-p_1)$ is positive 
and it acts like heat leakage term
which reduces the efficiency. On the other hand,
when $p_1>p_1'$, this term is negative which means that although each spin
locally
absorbs heat equal to $q_1$ from the hot bath, due to interaction the effective
total heat absorbed 
by the two-spin system is
less than $2 q_1$, which raises the efficiency for a given quantity of the work
performed.
It is interesting to know how much maximum gain in efficiency is possible for a 
given set of parameters .
We have proved an upper bound for the global efficiency, given by 
\begin {equation}
\eta  \le \frac{1-B_2/B_1}{1 -4J/B_1} < \eta_c,
\label{etabound}
\end {equation}
where $\eta_c = 1-T_2/T_1$ is the Carnot bound. 
Also for $\eta > \eta_0$, we have the condition $B_1 > 4 J$.
This implies that the ordering of energy levels which gives an 
enhancement of efficiency (over the uncoupled model) is:
\begin {equation}
 (2J - 2B_1)< -6J < 2J < (2J + 2B_1),
\end {equation}
and which after the first quantum adiabatic process, becomes
\begin {equation}
 (2J - 2B_2)< -6J < 2J < (2J + 2B_2).
\end {equation}
The proof of Eq. (\ref{etabound}) is given in the Appendix.
\section*{B. The case $B_2 > B_1$}
\label{<}
In this case, during the first quantum adiabatic process, the magnetic field
is \textit{increased} from its value $B_1$ to $B_2$.
If there is no interaction between the spins, the system cannot work as an
engine in
this case
because the condition $W>0$ will not be  satisfied  \cite{Kieu2004}. 
The conditions $T_1>T_2$  and $B_2>B_1$ directly lead to
\begin{eqnarray}
p_4&>&p_4',
\label{p4}\\
p_3&>&p_3'.
\label{p3}
\end{eqnarray}
Further, the positive work condition  implies $(p_2' - p_4') < (p_2- p_4)$,
which alongwith (\ref{p4}) gives
\begin{equation}
p_2>p_2'.
\label{p2}
\end{equation}
The normalisation of probabilities and the above three conditions Eqs. 
(\ref{p4}), (\ref{p3}) and (\ref{p2}) together give 
\begin{equation}
p_1'>p_1.
\label{p1}
\end{equation}
These are the necessary conditions for the system to work as an engine given
 that $T_1 > T_2$ and $B_2 > B_1$.
 According to Eq. (\ref{3}), the local work should be positive. 
This yields $q_1 < 0$ and $q_2>0$. This means locally the heat is absorbed
from the cold bath and given to the hot bath. 
Also the local efficiency is 
\begin{equation} 
\frac{w}{q_2}=1- \frac{B_1}{B_2}.
\end{equation}
Thus locally, the 
spins operate counter to the global 
temperature gradient present due to $T_1 > T_2$. But 
globally we do have $Q_1 > 0$ and $Q_2 < 0$. Thus the function of the two-spin
engine 
is consistent with the second law of thermodynamics, although locally
we seem to have a violation of the same. 
This apparent contradiction is resolved below using the concept of 
local effective temperatures.
\subsection*{C. Local temperatures}
Now each spin in the 2-spin system can be assigned a local effective
temperature,
corresponding to its local thermal state or the reduced density matrix
\cite{Garcia,Mahler2004,Hartmann}. 
This is true regardless of the state of the total system. Particularly, 
in stages 1 and 3 of the cycle, 
from Eqs. (\ref{rho}-\ref{rho_prime}) alongwith local Hamiltonian, we get the
local temperatures as
\begin{equation}
T_1'={2B_1}\;\left( {\log\!{\left[\frac{2}{(1+p_4-p_2)}-1\right]}}\right)^{-1}, 
\label{T}
\end{equation}
\begin{equation}
T_2'={2B_2}\; \left( {\log
\!{\left[\frac{2}{(1+p_4'-p_2')}-1\right]}}\right)^{-1}.
\label{T1}
\end{equation}
The important fact is that in the presence of interactions, the local
temperatures
are different from the corresponding bath temperatures. Thus $T_1' \ne T_1$ and
$T_2'\ne T_2$ if $J \ne 0$.
Further, since the work in our heat cycle is done only locally, 
the total work by the  system  can be regarded as equal to the work by two
independent spins operating  between their effective temperatures.

(i) {Engine working in $B_1>B_2$:}
 the positive work condition for a single spin is given by 
\begin{equation}
 \frac{B_2}{T_2'}> \frac{B_1} {T_1'}.
\end{equation} 
Since $B_1>B_2$, we get
\begin{equation}
 T_1'>T_2'.
\end{equation} 
At $J=0$, $T_1' = T_1$ and $T_2' = T_2$ and we have the result of Kieu's model
\cite{Kieu2004}.

(ii) {Engine working in $B_2>B_1$:}
in this model, the positive work condition is satisfied only when 
\begin{equation}
\frac{ B_1 }{T_1'} >\frac{ B_2}{T_2'}.
\label{b}
\end{equation}
Thus in this case  $T_2'>T_1'$.
Moreover, it can be shown from the definitions (\ref{T}) and (\ref{T1}) that 
for both the cases, $T_1'>T_1$ and  $T_2'>T_2$.
Finally, based on local temperatures, the counter-intuitive mechanism which
leads in 
case (ii) to $q_1<0$ and $q_2 >0$ can be justified as follows.
For $B_2>B_1$, due to the first adiabatic process, the local temperature
\textit{increases}
from $T_1'$ to  $T_1'(B_2/B_1)$. After contact with the cold bath, the local
temperature
becomes  $T_2'$, which due to condition (\ref{b}) is more than $T_1'(B_2/B_1)$. 
Thus heat should flow from the cold bath to the spin or $q_2 >0$. Similar
considerations lead to rejection of heat by the spin at the hot bath or $q_1<0$.
\section*{IV. Summary}
A model of coupled spins is used as working medium to realise a quantum Otto
engine.
The conditions for the 
efficiency to be higher than the non-interacting case are found. 
The antiferromagnetic interaction between the spins allows a fraction of the
total heat
to flow from cold bath to hot bath provided the total heat
should flow in in a direction suggested by global temperature gradient.
This mechanism increases the efficiency of the system compared to
non-interacting spins case.
 A tighter upper bound for the efficiency is found which is lower than the
Carnot value. 
The system can also work as a heat engine even if it undergoes an  adiabatic
compression 
($B_2>B_1$) in the second stage of the cycle. Here we have observed an
interesting 
mode of operation using the reduced density matrix whereby each spin absorbs
heat from  
the cold bath and rejects some heat to the hot bath while performing a net
work. 
This feature is also confirmed from the analysis of local effective temperatures of the
spins.
\section*{ACKNOWLEDGEMENT}
G.T. gratefully acknowledges financial support from Indian Institute of Science
Education
and Research Mohali.
\section*{APPENDIX}
\subsection*{Upper bound for global efficiency}
We consider the case of the engine working in the range $B_1>B_2$.
The condition to get a \textit{higher} efficiency as compared to uncoupled model
is the
case (ii)
discussed in Section IIIA and is given by
\begin {equation}
p_1>p_1'.
\label{p11}
\end {equation}
 From the condition $B_2 /T_2 > B_1 /T_1$ (Eq. (\ref{bt})), we get  
\begin{eqnarray}
 p_3&>&p_3',
\label{p33}\\
 p_4&>&p_4'.
\label{p44}
\end{eqnarray}
Then normalisation of the probabilities gives 
\begin {equation}
p_2'>p_2.
\label{p22}
\end {equation}
From Eqs. (\ref{p11}) and (\ref{p22}), we have
\begin {equation}
 \frac{p_2'}{p_1'}>\frac{p_2}{p_1},
\end {equation}
which simplifies to
\begin {equation}
 e^{(B_2-4J)/T_2}> e^{(B_1-4J)/T_1}.
\label{levels1}
\end {equation}
Fig. \ref{A1} shows  three possible ways of arranging the energy levels 
$(2J-2B_1)$ and  $-6J$ relative to  the level $(2J-2B_2)$  resulting from the 
first quantum adiabatic process. 
Equivalently, Eq. (\ref{levels1}) is of the form $e^x > e^y$, 
which may be satisfied in one of the following three ways: 
\begin{figure}[h]
\includegraphics{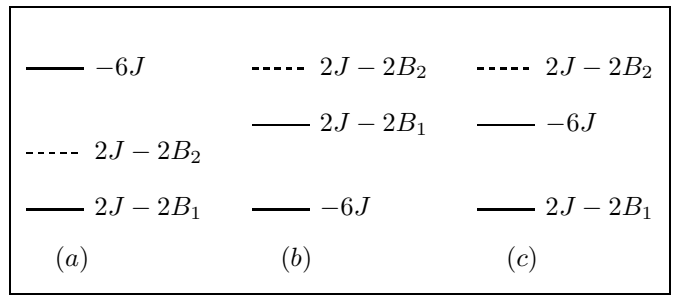}
\caption{Three possible configurations of energy levels with eigenvalues $-6J$,
$(2J-2B_1)$ 
and the level $(2J-2B_2)$ resulting from the first quantum adiabatic process
whereby  
$B_1$ is changed to a lower value $B_2$. Only case (a) is possible as discussed
in the 
Appendix.}
\label{A1}
\end{figure}
%

Case (a) represents $y>0$, $x>0$ and so $x>y$.  This implies, $B_1>4J$
and  $B_2>4J$. 

Case (b) represents $x<0$, $y<0$ and $|{x}| < |{y}|$. This implies
$B_1 < 4J$, $B_2 < 4J$, but due to the fact $T_2 /T_1 < 1$, we obtain 
$B_1 < B_2$ which leads to a contradiction.

Case (c) represents $y<0$ and $x>0$.  This possibility is also similarly ruled
out.

So the only possibility is case (a)
representing the fact that the energy levels $(2J-2B_1)$ and $(2J-2B_2)$ lie
below
the level $-6J$
when the coupled engine gives a \textit{higher} efficiency than the uncoupled
case.

When the inequality (\ref{levels1}) holds, we can write
\begin {equation}
 \frac{B_2-4J}{T_2}>\frac{B_1-4J}{T_1}.
\end {equation}
Since $B_1>4J$ , $B_2>4J$ and $T_1>T_2$, we get
\begin {equation}
\frac{\eta_0}{1-4J/B_1} < \eta_c = 1-\frac{T_2}{T_1},
\end {equation}
where $\eta_0 = 1-B_2/B_1$.
Now the global efficiency defined as $\eta = W/Q_1$, can be written as
\begin {equation}
 \eta=\frac{\eta_0}{1-\frac{4J(p_1-p_1')}{B_1(p_4-p_4'+p_2'-p_2)}}.
\end {equation}
From the inequalities between the probabilities (Eqs. (\ref{p11}),(\ref{p44})
and (\ref{p22})),
it follows that $(p_1-p_1')<(p_4-p_4'+p_2'-p_2)$.
Therefore, we finally obtain that when the efficiency is higher than the
uncoupled
case (or the lower bound is $\eta_0$), then an upper bound for efficiency is
given by
\begin {equation}
\eta < \frac{\eta_0}{1-4J/B_1}  < \eta_c.
\label{boundeta}
\end {equation}
When $J=0$, we have $\eta = \eta_0$.
A similar kind of proof can be constructed for the case $B_2>B_1$.
Interestingly,
the same bound as Eq. (\ref{boundeta}) is obtained.


\begin{references}
\bibitem{Scovil} H.E.D. Scovil and E.O. Schulz-Dubois, Phys. Rev. Lett. {\bf 2},
262 (1959); J.E. Geusic, E.O. Schulz-Dubois, and H.E.D. Scovil, Phys. Rev. {\bf
156}, 343 (1967)
\bibitem{Kieu2004} T.D. Kieu, Phys. Rev. Lett. {\bf 93}, 140403 (2004);
Eur. Phys. J. D {\bf 39}, 115 (2006).
%
\bibitem{TZhang} T. Zhang, W.-T. Liu, P.-X. Chen, and C.-Z. Li, Phys. Rev. A
{\bf 75}, 062102 (2007).
%
\bibitem{GFZhang}G. F. Zhang, Eur. Phys. J. D {\bf 49}, 123 (2008).
%
\bibitem{Wang2009} Hao Wang, Sanqiu Liu, and Jizhou He, Phys. Rev. E {\bf 79},
041113 (2009).
%
\bibitem{Scully}M.O. Scully, M.S. Zubairy, G.S. Agarwal, and H. Walther,
Science {\bf 299}, 862 (2003).
%
\bibitem{Lutz}R. Dillenschneider and E. Lutz, Europhys.Lett. {\bf 88}, 5003
(2009).
%
\bibitem{Leff} {\it Maxwell's Demon 2: Entropy, Classical and Quantum
Information, Computing}, 
edited by H.S. Leff and A.F. Rex (Institute of Physics, Bristol, 2003).
%
\bibitem{Quan2006}H. T. Quan, Y. D. Wang, Y. Liu, C.P. Sun, and F. Nori, Phys.
Rev. Lett. {\bf 97}, 180402 (2006).
%
\bibitem{Zhou2010}Yun Zhou and Dvira Segal, Phys. Rev. E {\bf 82}, 011120
(2010).
%
\bibitem{Allahv2010} A.E. Allahverdyan, K. Hovhannisyan and G. Mahler,
Phys. Rev. E \textbf{81}, 051129 (2010). 
\bibitem{Noah2010}Noah Linden, Sandu Popescu, and Paul Skrzypczyk,  Phys. Rev.
Lett. {\bf105}, 130401 (2010).
%
\bibitem{Paul}Paul Skrzypczyk, Nicolas Brunner, Noah Linden, and Sandu Popescu,
arXiv:quant-ph/1009.0865.
%

%
\bibitem{Quan2007}H. T. Quan,  Yu-xi Liu, C.P. Sun, and F. Nori, Phys.
Rev. E {\bf 76}, 031105 (2007).
%
\bibitem{Humphrey2002}T.E. Humphrey, R.Newbury, R.P Taylor, and H. Linke, Phys.
Rev. Lett. {\bf
89}, 116801 (2002).
%
\bibitem{Geva1992} E. Geva and R. Kosloff, J. Chem. Phys. {\bf 96}, 3054 (1992).
%
\bibitem{Broeck2005} C. Van den Broeck, Phys. Rev. Lett. {\bf 95}, 190602
(2005).
%
\bibitem{Feng} Feng Wu, Lingen Chen, Shuang Wu, Fengrui Sun and Chih Wu, J.
Chem. Phys. {\bf 124}, 21472 (2006).
%
\bibitem{AJM2008} A. E. Allahverdyan, R. S. Johal, and G. Mahler, Phys. Rev. E
{\bf 77}, 041118 (2008).
%
\bibitem{Esposito2009} M. Esposito, K. Lindenberg, and C. Van den Broeck,
 Phys. Rev. Lett. \textbf{102}, 130602 (2009).
%
\bibitem{ELB2009} M. Esposito, K. Lindenberg, and C. Van den Broeck, Europhys.
Lett. \textbf{85}, 60010 (2009).
\bibitem{EKLB2010} M. Esposito, R. Kawai, K. Lindenberg, and C. Van den Broeck,
Phys. Rev. E \textbf{81}, 041106 (2010). 
%
\bibitem{Arnesen} M.C. Arnesen, S. Bose, and V. Vedral, Phys. Rev. Lett.
{\bf87}, 017901 (2001)
%
\bibitem{Garcia}A. Garcia-Saez, A. Ferraro, and A. Acin, Phys. Rev. A {\bf 79},
052340 (2009).
%
\bibitem{Mahler2004} J. Gemmer, M. Michel, and G. Mahler, {\it Quantum
Thermodynamics}, Springer, Berlin (2004).
%
\bibitem{Hartmann} M. Hartmann, Contemp. Phys. {\bf 47}, 89 (2006).
%
\end{references}
\end{document}